# Comparison of micromagnetic parameters of ferromagnetic semiconductors (Ga,Mn)(As,P) and (Ga,Mn)As

N. Tesařová[1], D. Butkovičová[1], R. P. Campion[2], A.W. Rushforth[2], K. W. Edmonds[2], P. Wadley[2], B. L. Gallagher[2], E. Schmoranzerová,[1] F. Trojánek[1], P. Malý[1], P. Motloch[4], V. Novák[3], T. Jungwirth[3,2], and P. Němec[1,*]

[1] Faculty of Mathematics and Physics, Charles University in Prague, Ke Karlovu 3, 121 16 Prague 2, Czech Republic
[2] School of Physics and Astronomy, University of Nottingham, Nottingham NG72RD, United Kingdom
[3] Institute of Physics ASCR, v.v.i., Cukrovarnická 10, 16253 Prague 6, Czech Republic
[4] University of Chicago, Chicago, IL 60637, USA

We report on the determination of micromagnetic parameters of epilayers of the ferromagnetic semiconductor (Ga,Mn)As, which has easy axis in the sample plane, and (Ga,Mn)(As,P) which has easy axis perpendicular to the sample plane. We use an optical analog of ferromagnetic resonance where the laser-pulse-induced precession of magnetization is measured directly in the time domain. By the analysis of a single set of pump-and-probe magneto-optical data we determined the magnetic anisotropy fields, the spin stiffness and the Gilbert damping constant in these two materials. We show that incorporation of 10% of phosphorus in (Ga,Mn)As with 6% of manganese leads not only to the expected sign change of the perpendicular-to-plane anisotropy field but also to an increase of the Gilbert damping and to a reduction of the spin stiffness. The observed changes in the micromagnetic parameters upon incorporating P in (Ga,Mn)As are consistent with the reduced hole density, conductivity, and Curie temperature of the (Ga,Mn)(As,P) material. We report that the magnetization precession damping is stronger for the $n = 1$ spin wave resonance mode than for the $n = 0$ uniform magnetization precession mode.

## I. INTRODUCTION

(Ga,Mn)As is the most widely studied diluted magnetic semiconductor (DMS) with a carrier-mediated ferromagnetism.[1] Investigation of this material system can provide fundamental insight into new physical phenomena that are present also in other types of magnetic materials – like ferromagnetic metals – where they can be exploited in spintronic applications.[2-5] Moreover, the carrier concentration in DMSs is several orders of magnitude lower than in conventional FM metals which enables manipulation of magnetization by external stimuli – e.g. by electric[6,7] and optical[8,9] fields. Another remarkable property of this material is a strong sensitivity of the magnetic anisotropy to the epitaxial strain. (Ga,Mn)As epilayers are usually prepared on a GaAs substrate where the growth-induced compressive strain leads to in-plane orientation of the easy axis (EA) for Mn concentrations ≥2%.[10] However, for certain experiments – e.g., for a visualization of magnetization orientation by the magneto-optical polar Kerr effect[11-17] or the anomalous Hall effect[12,18] – the EA orientation in the direction perpendicular to the sample plane is more suitable. To achieve this, (Ga,Mn)As layers have been grown on relaxed (In,Ga)As buffer layers that introduce a tensile strain in (Ga,Mn)As.[11,12,14,16-18] However, the growth on (In,Ga)As layers can result in a high density of line defects that can lead to high coercivities and a strong pinning of domain walls (DW).[16,17] Alternatively, tensile strain and perpendicular-to-plane orientation of the EA can be achieved by incorporation of small amounts of phosphorus in (Ga,Mn)(As,P) layers.[19,20] In these epilayers, the EA can be in the sample plane for the as-grown material and perpendicular to the plane for fully annealed (Ga,Mn)(As,P).[21] The possibility of magnetic anisotropy fine tuning by the thermal annealing turns out to be a very favorable property of (Ga,Mn)(As,P) because it enables the preparation of materials with extremely low barriers for magnetization switching.[22,23] Compared to tensile-stained (Ga,Mn)As/(In,Ga)As films, (Ga,Mn)(As,P)/GaAs epilayers show weaker DW pinning, which allows observation of the intrinsic flow regimes of DW propagation.[13,15,24]

Preparation of uniform (Ga,Mn)As epilayers with minimized density of unintentional extrinsic defects is a rather challenging task which requires optimized growth and post-growth annealing conditions.[25] Moreover, the subsequent determination of material micromagnetic parameters by the standard characterization techniques, such as ferromagnetic resonance (FMR), is complicated by the fact that these techniques require rather thick films, which may be magnetically inhomogeneous.[25,26] Recently, we have reported the preparation of high-quality (Ga,Mn)As epilayers where the individually optimized synthesis protocols yielded systematic doping trends, which are microscopically well understood.[25] Simultaneously with the optimization of the material synthesis, we developed an optical analog of FMR (optical-FMR)[25], where all micromagnetic parameters of the in-plane (Ga,Mn)As were deduced from a single magneto-optical (MO) pump-and-probe experiment where a laser pulse induces precession of magnetization.[27,28] In this method the anisotropy fields are determined from the dependence of the precession frequency on the magnitude and the orientation of the external magnetic field, the Gilbert damping constant is deduced from the damping of the precession signal, and the spin stiffness is obtained from the mutual spacing of the spin wave resonance modes observed in the measured MO signal. In this paper we apply this all optical-FMR to (Ga,Mn)(As,P). We demonstrate the applicability of this method also for the determination of micromagnetic parameters in DMS materials with a perpendicular-to-plane orientation of the EA. By this method we show that the incorporation of P in (Ga,Mn)As leads not only to the expected sign change of the perpendicular-to-plane anisotropy field but also to a considerable increase of the Gilbert damping and to a reduction of the spin stiffness. Moreover, we illustrate that the all optical-FMR can be very effectively used not only for an investigation of the uniform magnetization precession but also for a study of spin wave resonances.

## II. EXPERIMENTAL

In our previous work we reported in detail on the preparation and micromagnetic characterization of (Ga,Mn)As epilayers prepared in MBE laboratory in Prague.[25] We also pointed out that the preparation of (Ga,Mn)As by this highly non-equilibrium synthesis in two distinct MBE laboratories in Prague and in Nottingham led to a growth of epilayers with micromagnetic parameters that showed the same doping trends.[25] Nevertheless, the preparation of epilayers with identical parameters (e.g., thickness, nominal Mn content, etc.) in two distinct MBE machines is still a nontrivial task. Therefore, in this study of the role of the phosphorus incorporation to (Ga,Mn)As we opted for a direct comparison of materials prepared in one MBE machine. The investigated $Ga_{1-x}Mn_xAs$ and $Ga_{1-x}Mn_xAs_{1-y}P_y$ epilayers were prepared in Nottingham[20] with the same nominal amount of Mn ($x = 6\%$) and the same growth time on a GaAs substrate (with 50 nm thick GaAsP buffer layer in the case of (Ga,Mn)(As,P)]. They differ only in the incorporation of P ($y = 10\%$) in the latter epilayer.

The inferred epilayer thicknesses are (24.5 ± 1.0) nm for both (Ga,Mn)As and (Ga,Mn)(As,P).[29] The as-grown layers, which both had the EA in the epilayer plane, were thermally annealed (for 48 hours at 180°C). This led to an increase in Curie temperature and to a rotation of the EA to the perpendicular-to-plane orientation for (Ga,Mn)(As,P).[20,21]

The magnetic anisotropy of the samples was studied using a superconducting quantum interference device (SQUID) magnetometer and by the all-optical FMR.[25] The hole concentration was determined by fitting to Hall effect measurements at low temperatures (1.8 K) for external magnetic fields from 2 T to 6 T. In this range the magnetization is saturated and one can obtain the normal Hall coefficient after correction for the field dependence of the anomalous Hall due to the weak magnetoresistance.[30] The time-resolved pump-and-probe MO experiments were performed using a titanium sapphire pulsed laser (pulse width ≈ 200 fs) with a repetition rate of 82 MHz, which was tuned ($h\nu$ = 1.64 eV) above the GaAs band gap. The energy fluence of the pump pulses was around 30 $\mu Jcm^{-2}$ and the probe pulses were at least ten times weaker. The pump pulses were circularly polarized (with a helicity controlled by a quarter wave plate) and the probe pulses were linearly polarized (in a direction perpendicular to the external magnetic field). The time-resolved MO data reported here correspond to the polarization-independent part of the pump-induced rotation of probe polarization plane, which was computed from the measured data by averaging the signals obtained for the opposite helicities of circularly polarized pump pulses.[27,28] The experiment was performed close to the normal-incidence geometry, where the angles of incidence were 9° and 3° (measured from the sample normal) for the probe and the pump pulses, respectively.

The rotation of the probe polarization plane is caused by two MO effects – the polar Kerr effect and the magnetic linear dichroism, which are sensitive to perpendicular-to-plane and in-plane components of magnetization, respectively.[31-33] For all MO experiments, samples were mounted in a cryostat and cooled down to ≈ 15 K. The cryostat was placed between the poles of an electromagnet and the external magnetic field $H_{ext}$ ranging from ≈ 0 to 585 mT was applied in the sample plane, either in the [010] or [110] crystallographic direction of the sample (see inset in Fig. 1 for a definition of the coordinate system). Prior to all measurements, we always prepared the magnetization in a well-defined state by first applying a strong saturating magnetic field and then reducing it to the desired magnitude of $H_{ext}$.

### III. RESULTS AND DISCUSSION

#### A. Sample characterization

The hysteresis loops measured by SQUID magnetometry for external magnetic field applied along the in-plane [-110] and perpendicular-to-plane [001] crystallographic directions in (Ga,Mn)As and (Ga,Mn)(As,P) samples are shown in Fig. 1(a) and Fig. 1(c), respectively. These data confirm the expected in-plane and perpendicular-to-plane orientations of the EA in (Ga,Mn)As and (Ga,Mn)(As,P), respectively. Moreover, they reveal that for the (Ga,Mn)(As,P) sample, an external magnetic field of ≈ 250 mT is needed to rotate the magnetization into the sample plane. In Fig. 1(b) and Fig. 1(d) we show the temperature dependences of the remanent magnetization of the samples from which the Curie temperature $T_c$ of ≈ 130 K and ≈ 110 K can be deduced. The measured saturation magnetization also indicates very similar density of Mn moments contributing to the ferromagnetic state in the two samples.

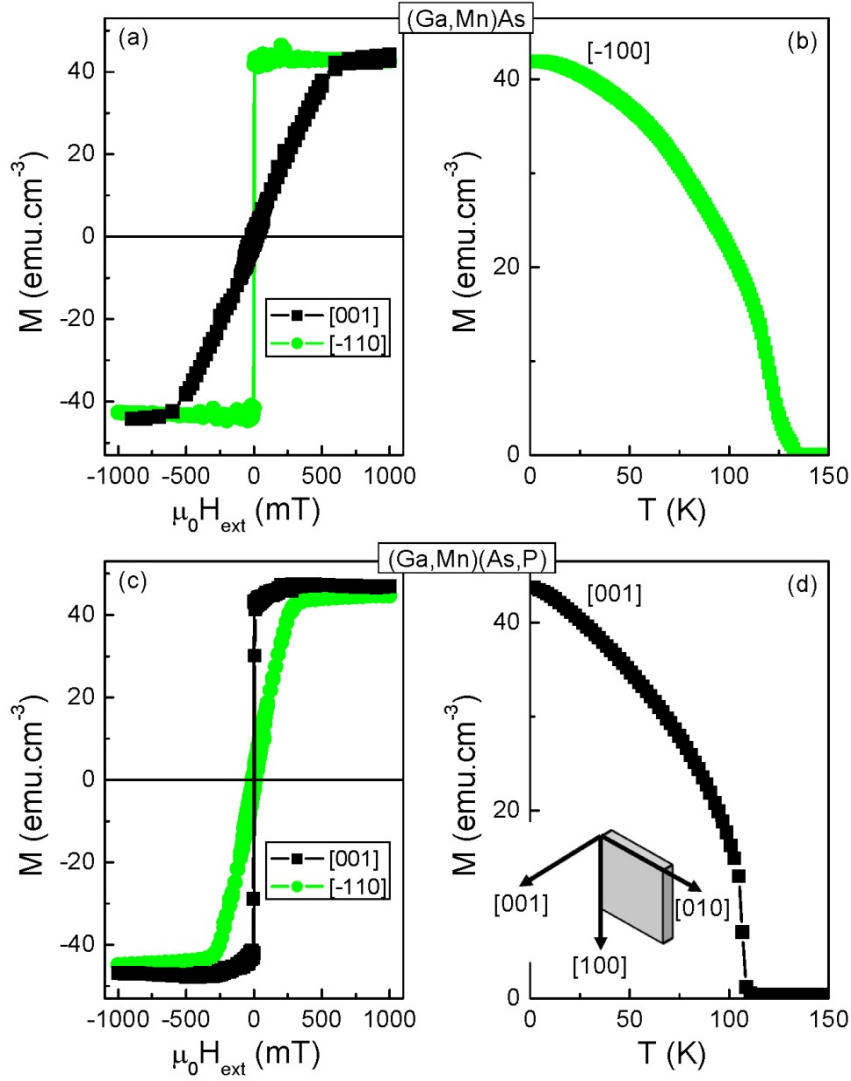

Fig. 1 (Color online): Magnetic characterization of samples: (a), (b) (Ga,Mn)As and (c), (d) (Ga,Mn)(As,P). (a), (c) Hysteresis loops measured in at 2 K for the external magnetic field applied in the sample plane (along the crystallographic direction [-110]) and perpendicular to sample plane (along the crystallographic direction [001]). (b), (d) Temperature dependence of the remanent magnetization. Inset: Definition of the coordinate system.

The electrical characterization of the samples is shown in Fig. 2. The measured data show a sharp Curie point singularity in the temperature derivative of the resistivity which confirms the high quality of the samples.[25] The hole densities inferred from Hall measurements are $(1.3 \pm 0.2) \times 10^{21}$ cm$^{-3}$ and $(0.8 \pm 0.2) \times 10^{21}$ cm$^{-3}$ for (Ga,Mn)As and (Ga,Mn)(As,P), respectively. The hole density obtained for (Ga,Mn)As is in agreement with our previous measurements for similar films in magnetic fields up 14 T.[30] The reduction of the density of itinerant holes quantitatively correlates with the observed increase of the resistivity of the (Ga,Mn)(As,P) film as compared to the (Ga,Mn)As sample.

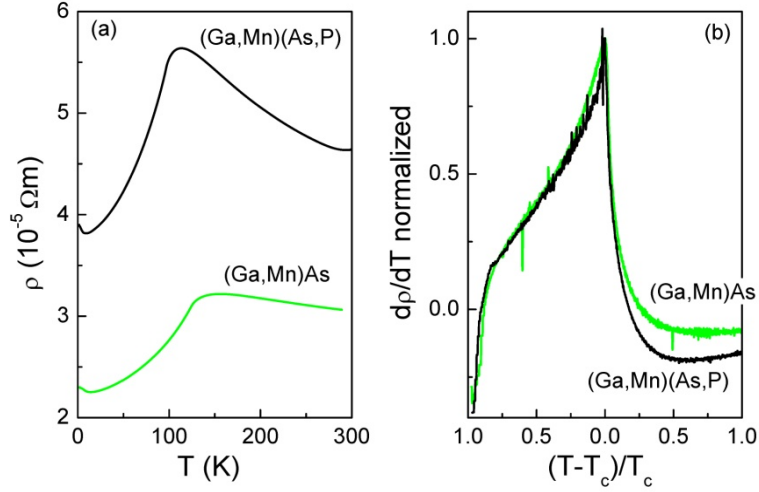

Fig. 2 (Color online): Electrical characterization of samples. Temperature dependence of the resistivity (a) and its temperature derivative (b).

**B. Time-resolved magneto-optical experiment**

In Fig. 3(a) and 3(b) we show the measured MO signals that reflect the magnetization dynamics in (Ga,Mn)As and (Ga,Mn)(As,P) samples, respectively. These signals can be decomposed into the oscillatory parts [Figs. 3(c) and 3(d)] and the non-oscillatory pulse-like background [Fig. 3(e) and 3(f)].[27, 28] The oscillatory part arises from the precessional motion of magnetization around the quasi-equilibrium EA and the pulse-like function reflects the laser-induced tilt of the EA and the laser-induced demagnetization.[25,31] The pump polarization-independent MO data reported here, which were measured at a relatively low excitation intensity of 30 $\mu$Jcm$^{-2}$, can be attributed to the magnetization precession induced by a transient heating of the sample due to the absorption of the laser pulse.[8,9] Before absorption of the pump pulse the magnetization is along the EA direction. Absorption of the laser pulse leads to a photo-injection of electron-hole pairs. The subsequent fast non-radiative recombination of photo-injected electrons induces a transient increase of the lattice temperature (within tens of picoseconds after the impact of the pump pulse). The laser-induced change of the lattice temperature then leads to a change of the EA position.[34] As a result, magnetization starts to follow the EA shift by the precessional motion. Finally, dissipation of the heat leads to a return of the EA to the equilibrium position and the precession of magnetization is stopped by a Gilbert damping.[25] It is apparent from Fig. 3 that the measured MO signals are strongly dependent on a magnitude of the external magnetic field, which was applied in the epilayer plane along the [010] crystallographic direction in both samples. In particular, absorption of the laser pulse does not induce precession of magnetization in (Ga,Mn)(As,P) unless magnetic field stronger than 20 mT is applied [see Fig. 3(d)].

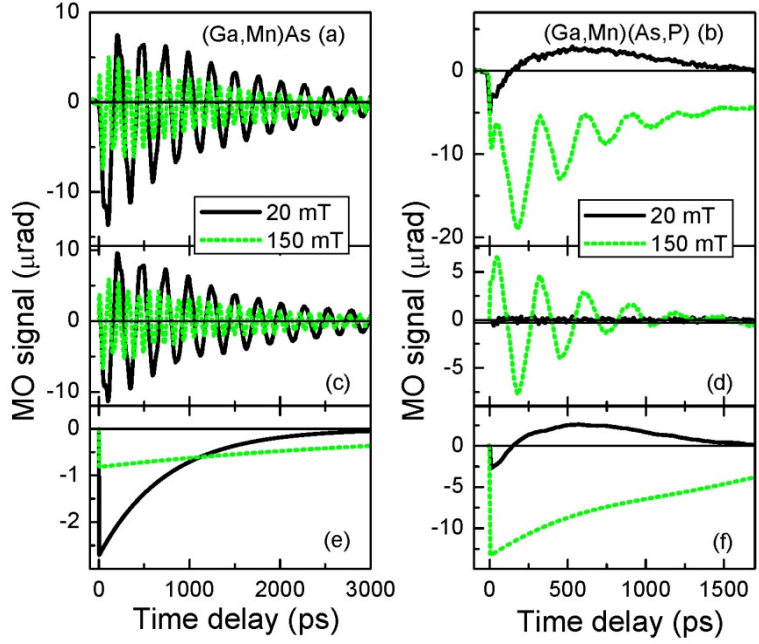

Fig. 3 (Color online): Time-resolved magneto-optical (MO) signals measured in (Ga,Mn)As (a) and (Ga,Mn)(As,P) (b) for two magnitudes of the external magnetic field applied along the [010] crystallographic direction. The measured MO signals were decomposed into oscillatory parts [(c) and (d)], which correspond to the magnetization precession, and to non-oscillatory parts [(e) and (f)], which are connected with the quasi-equilibrium tilt of the easy axis and with the demagnetization. Note different *x*-scales in the left and in the right columns.

The magnetization dynamics is described by the Landau-Lifshitz-Gilbert (LLG) equation that is usually expressed in the form[35,36]:

$$\frac{d\boldsymbol{M}(t)}{dt} = -\gamma[\boldsymbol{M}(t) \times \boldsymbol{H}_{eff}(t)] + \frac{\alpha}{M_s}\left[\boldsymbol{M}(t) \times \frac{d\boldsymbol{M}(t)}{dt}\right], \quad (1)$$

where $\gamma = (g\mu_B)/\hbar$ is the gyromagnetic ratio, $g$ is the Landé g-factor, $\mu_B$ is the Bohr magneton, $\hbar$ is the reduced Planck constant, $\alpha$ is the Gilbert damping constant, and $H_{eff}$ is the effective magnetic field. Nevertheless, it is more convenient to express this equation in spherical coordinates where the direction of the magnetization vector $M$ is given by the polar angle $\theta$ and azimuthal angle $\varphi$ and where $H_{eff}$ can be directly connected with angular derivatives of the free energy density functional $F$ (see the Appendix).[37] For small deviations $\delta\theta$ and $\delta\varphi$ of magnetization from its equilibrium position (given by $\theta_0$ and $\varphi_0$), the solution of LLG equation can be written in the form $\theta(t) = \theta_0 + \delta\theta(t)$ and $\varphi(t) = \varphi_0 + \delta\varphi(t)$ as

$$\theta(t) = \theta_0 + A_\theta e^{-k_d t}\cos(2\pi f t + \Phi_\theta), \quad (2)$$
$$\varphi(t) = \varphi_0 + A_\varphi e^{-k_d t}\cos(2\pi f t + \Phi_\varphi), \quad (3)$$

where the constants $A_\theta$ ($A_\varphi$) and $\Phi_\theta$ ($\Phi_\varphi$) represent the initial amplitude and phase of $\theta$ ($\varphi$), respectively, $f$ is the magnetization precession frequency, and $k_d$ is the precession damping rate (see the Appendix). The precession frequency reflects the internal magnetic anisotropy of the sample that can be characterized by the cubic ($K_C$), in-plane uniaxial ($K_u$) and out-of-plane uniaxial ($K_{out}$) anisotropy fields (see Eq. (A4) in the Appendix).[10] Moreover, $f$ depends also on the magnitude and on the orientation of $H_{ext}$ (see the Appendix) and, therefore, the magnetic field dependence of $f$ can be used to evaluate the magnetic anisotropy fields in the sample. If the applied in-plane magnetic field is strong enough to align the magnetization parallel with $H_{ext}$ (i.e., for $H_{ext}$ exceeding the saturation field in the sample for a particular orientation of $H_{ext}$), $\theta = \theta_H = \pi/2$ and $\varphi = \varphi_H$ and if the precession damping is relatively slow, i.e. $\alpha^2 \approx 0$ $f$ can be expressed as

$$f = \frac{g\mu_B}{h}\sqrt{\begin{pmatrix}H_{ext} - 2K_{out} + \frac{K_C(3+cos4\varphi)}{2} + 2K_u sin^2\left(\varphi_H - \frac{\pi}{4}\right)\end{pmatrix} \times (H_{ext} + 2K_C cos4\varphi_H - 2K_u sin2\varphi_H)}, \quad (4)$$

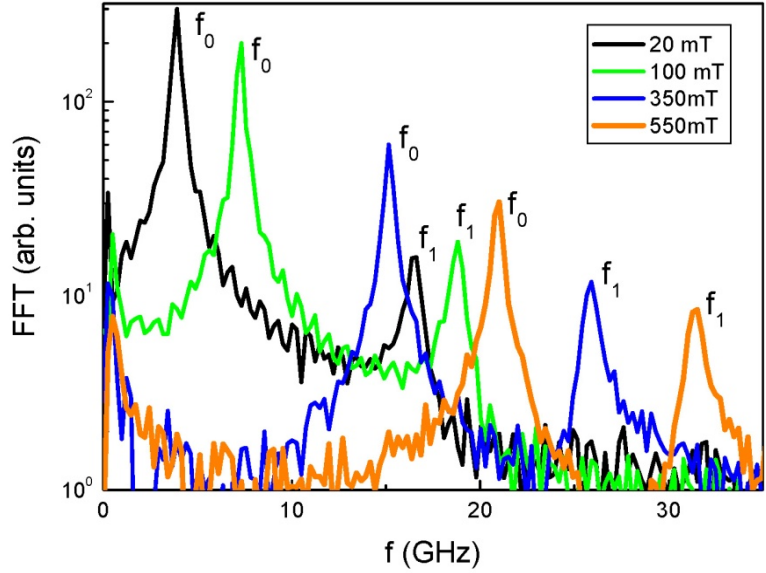

Fig. 4 (Color online): Fourier spectrum of the oscillatory part of the MO signal measured in (Ga,Mn)As for external magnetic fields applied along the [010] crystallographic direction. $f_0$ and $f_1$ indicate the frequencies of the uniform magnetization precession and the first spin wave resonance, respectively.

In Fig. 4 we show the fast Fourier transform (FFT) spectra of the oscillatory parts of the MO signals measured in the (Ga,Mn)As sample for different values of $H_{ext}$. This figure clearly reveals that for all external magnetic fields there are two distinct oscillatory frequencies present in the measured data. These precession modes are the spin wave resonances (SWRs) – i.e., spin waves (or magnons) that are selectively amplified by fulfilling the boundary conditions: In a homogeneous thin magnetic film with a thickness $L$, only the perpendicular standing waves with a wave vector $k$ fulfilling the resonant condition $kL = n\pi$ (where $n$ is the mode number) are amplified.[25,38-41] In our case – using the ferromagnetic films with a thickness around 25 nm – we detect only[42] the uniform magnetization precession with zero $k$ vector (i.e. the precession where at any instant of time all magnetic moments are parallel over the entire sample; $n = 0$ at frequency $f_0$) and the first SWR (i.e. $n = 1$ at frequency $f_1$). See the inset in Fig. 8 for a schematic depiction of the modes. In Fig. 5 we plot the amplitudes of the uniform magnetization precession ($A_0$) and of the first SWR ($A_1$) as a function of the external magnetic field $H_{ext}$. In the (Ga,Mn)As sample, the oscillations are present even when no magnetic field is applied and the precession amplitude increases slightly with an increasing $H_{ext}$ (up to $\approx 20$ mT for $A_0$ and up to $\approx 60$ mT for $A_1$). Above this value, a further increase of $H_{ext}$ leads to a suppression of the oscillations, but the suppression of the first SWR is slower than that of the uniform magnetization precession [see Fig. 5(c)]. In (Ga,Mn)(As,P), the oscillatory signal starts to appear at ≈ 50 mT, reaches its maximum for $H_{ext}$ ≈ 175 mT, and a further increase of $H_{ext}$ leads to its monotonic decrease, like in the case of (Ga,Mn)As. The observed field dependence of the precession amplitude, which expresses the sensitivity of the EA position on the laser-induced sample temperature change, can be qualitatively understood as follows. In (Ga,Mn)As, the position of the EA in the sample plane is given by a competition between the cubic and the in-plane uniaxial magnetic anisotropies.[10,25] The laser-induced heating of the sample leads to a reduction of the magnetization magnitude $M$ and, consequently, it enhances the uniaxial anisotropy relative to the cubic anisotropy.[9] This is because the uniaxial anisotropy component scales with magnetization as ~ $M^2$ while the cubic component scales as ~ $M^4$. The application of $H_{ext}$ along the [010] crystallographic direction deepens the minimum in the [010] direction in the free energy density functional $F$ (due to the Zeeman term in $F$, see Eq. (A4) in the Appendix). Measured data shown in Fig. 5 reveal that in the (Ga,Mn)As sample, $H_{ext}$ initially (for $H_{ext}$ up to ≈ 20 mT) destabilizes the position of EA but stabilizes it for large values of $H_{ext}$ (where the position of the energy minimum in $F$ is dominated by the Zeeman term, which is not temperature dependent). In the case of (Ga,Mn)(As,P), the position of the EA is determined by the strong perpendicular-to-plane anisotropy. Therefore, without an external magnetic field, the laser-induced heating of the sample does not change significantly the position of EA and, consequently, does not initiate the precession of magnetization [see Fig. 5(b)]. The application of an in-plane field moves the energy minimum in $F$ towards the sample plane [see Fig. 1(c)] which makes the EA position more sensitive to the laser-induced temperature change. Finally, for a sufficiently strong $H_{ext}$, the sample magnetic anisotropy is dominated by the temperature-independent Zeeman term, which again suppresses the precession amplitude. The markedly different ratio $A_1/A_0$ in the (Ga,Mn)As and (Ga,Mn)(As,P) samples is probably connected with a different surface magnetic anisotropy and/or a slight difference in magnetic homogeneity in these two samples.[43,44]

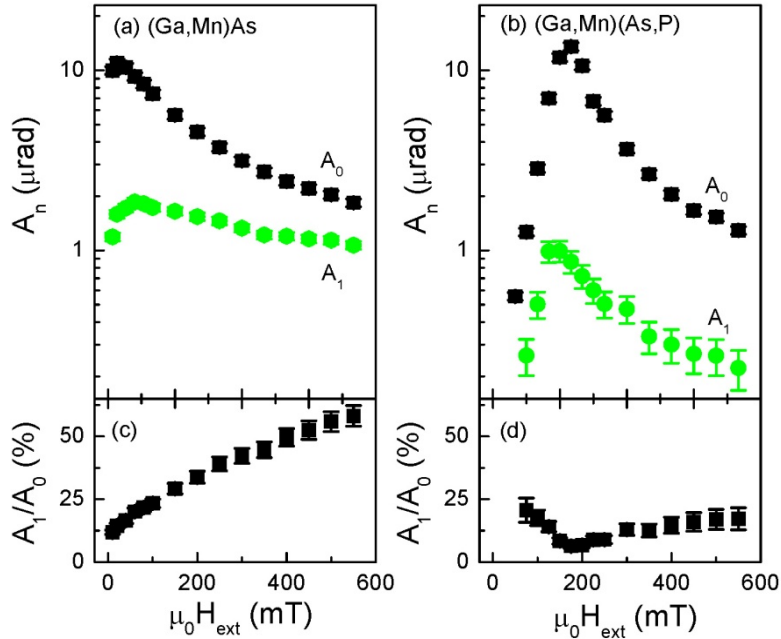

Fig. 5 (Color online): Dependence of the amplitude of the uniform magnetization precession ($A_0$) and the first spin wave resonance ($A_1$) on the magnitude of the external magnetic field ($H_{ext}$) applied along the [010] crystallographic direction in (Ga,Mn)As (a) and (Ga,Mn)(As,P) (b). (c) and (d) Dependence of the ratio $A_1 / A_0$ on $H_{ext}$.

## C. Determination of magnetic anisotropy

In Fig. 6 we plot the magnetic field dependences of $f_0$ and $f_1$ for two different orientations of $H_{ext}$. The frequency $f_0$ of the spatially uniform precession of magnetization is given by Eq. (4). For the SWRs, where the local moments are no longer parallel (see the inset in Fig. 8), restoring torques due to exchange interaction and internal magnetic dipolar interaction have to be included in the analysis.[39-41,45] For $H_{ext}$ along the [010] crystallographic direction (i.e., for $\varphi_H = \pi/2$) Eq. (4) can be written as

$$f_n = \frac{g\mu_B}{h}\sqrt{(H_{ext} - 2K_{out} + K_c + \Delta H_n)(H_{ext} - 2K_c - 2K_u + \Delta H_n)}, \quad (5)$$

where $\Delta H_n$ is the shift of the resonant field for the $n^{th}$ spin-wave mode with respect to the $n = 0$ uniform precession mode. Analogically, for $H_{ext}$ applied in the [110] crystallographic direction (i.e., for $\varphi_H = \pi/4$)

$$f_n = \frac{g\mu_B}{h}\sqrt{(H_{ext} - 2K_{out} + 2K_c + K_u + \Delta H_n)(H_{ext} + 2K_c + \Delta H_n)}. \quad (6)$$

The lines in Fig. 6 represent the fits of all four measured dependencies $f_n = f_n (H_{ext}, \varphi_H)$ [where $n = 0; 1$ and $\varphi_H = \pi/4; \pi/2$] with a single set of anisotropy constants for each of the samples, which confirms the credibility of the fitting procedure. The obtained anisotropy constants at ≈ 15 K are: $K_C = (17 \pm 3)$ mT, $K_u = (11 \pm 5)$ mT, $K_{out} = (-200 \pm 20)$ mT for (Ga,Mn)As and $K_C = (14 \pm 3)$ mT, $K_u = (11 \pm 5)$ mT, $K_{out} = (90 \pm 10)$ mT for (Ga,Mn)(As,P), respectively (in both cases we considered the Mn g-factor of 2). For (Ga,Mn)As, we can now compare these anisotropy constants with those obtained by the same fitting procedure for samples prepared in a different MBE laboratory (in Prague) – see Fig. 4 in Ref. 25. We see that the previously reported[25] doping trends of $K_C$ and $K_{out}$ predict for a sample with nominal Mn doping $x = 6\%$ the anisotropy fields which are the same as those reported in this paper for the sample grown in Nottingham. This observation is in accord with the current microscopic understanding of their origin – $K_C$ reflects the zinc-blende crystal structure of the host semiconductor and $K_{out}$ is a sum of the anisotropy due to the growth-induced lattice-matching strain and of the thin-film shape anisotropy, which should be the same for equally doped and optimally synthesized samples, independent of the growth chamber. On the other hand, the microscopic origin of in-plane uniaxial anisotropy field $K_u$ is still not established[10,25] and our data reveal that it is considerably smaller in the sample grown in Nottingham. The incorporation of phosphorus does not change significantly the values of $K_C$ and $K_u$ but it strongly modifies the magnitude and changes the sign of $K_{out}$, which is in agreement with the previous results obtained by FMR experiment.[22]

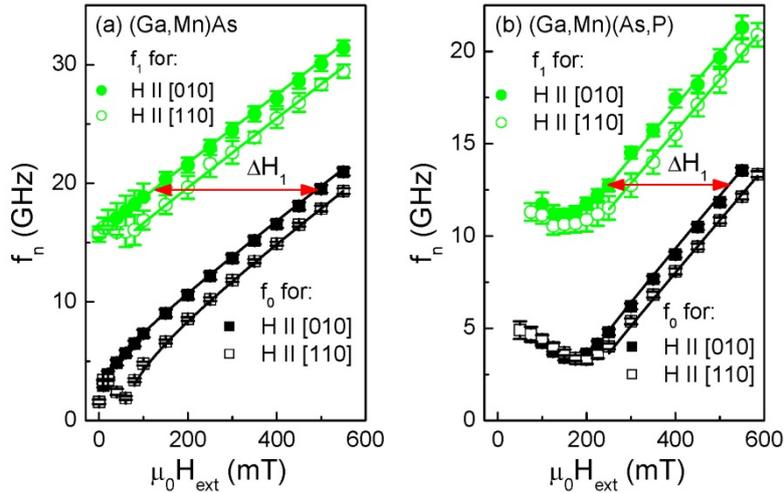

Fig. 6 (Color online): Magnetic field dependence of the precession frequencies $f_0$ and $f_1$ for two different orientations of the external magnetic field (points) measured in (Ga,Mn)As (a) and (Ga,Mn)(As,P) (b). Lines are the fits by Eqs. (5) and (6). $\Delta H_1$ indicates the shift of the resonant field for the first spin-wave mode with respect to the uniform precession mode.

### D. Determination of spin stiffness

The observation of a higher-order SWR enables us to also determine the exchange spin stiffness constant $D$, which is a parameter that is rather difficult to extract from other experiments in (Ga,Mn)As.[25,46] In homogeneous thin films, $\Delta H_n$ is given by the Kittel formula[43]

$$\Delta H_n \equiv H_0 - H_n = n^2 \frac{D}{g\mu_B} \frac{\pi^2}{L^2}, \qquad (7)$$

where $L$ is the thickness of the magnetic film. By fitting the data in Fig. 6, we obtained $\Delta H_1 = (363 \pm 2)$ mT for (Ga,Mn)As and $(271 \pm 2)$ mT for (Ga,Mn)(As,P) which correspond to $D = (2.5 \pm 0.2)$ meVnm$^2$ and $(1.9 \pm 0.2)$ meVnm$^2$ for (Ga,Mn)As and (Ga,Mn)(As,P), respectively (note that the relatively large experimental error in $D$ is given mainly by the uncertainty of the epilayer thickness).[29] The value obtained for (Ga,Mn)As is again in agreement with that reported previously for samples grown in Prague,[25] which also confirms the consistent determination of the epilayer thicknesses in both MBE laboratories.[29] The incorporation of phosphorus leads to a reduction of $D$ which correlates with the decrease of the hole density,[47] and the reduced $T_c$ in (Ga,Mn)(As,P), as compared to its (Ga,Mn)As counterpart.

### E. Determination of Gilbert damping

The Gilbert damping constant $\alpha$ can be determined by fitting the measured dynamical MO signals by the LLG equation.[35,36,48] For a relatively slow precession damping and a sufficiently strong external magnetic field, the analytical solution of the LLG equation gives (see the Appendix)

$$k_d = \alpha \frac{g\mu_B}{2\hbar} \left( 2H_{ext} - 2K_{out} + \frac{K_C(3+5cos4\varphi_H)}{2} + K_u(1 - 3sin2\varphi_H) \right). \qquad (8)$$

Eq. (8) shows not only that $k_d$ is proportional to $\alpha$ but also that for obtaining a correct value of $\alpha$ from the measured MO precession signal damping it is necessary to take into account a realistic magnetic anisotropy of the investigated sample. Nevertheless, the correct dependence of $k_d$ on magnetic anisotropy was not considered in the previous studies[35,36,48] where only one effective magnetic field was used, which is probably one of the reasons why mutually inconsistent results were obtained for Ga$_{1-x}$Mn$_x$As with a different Mn content $x$. An increase of $\alpha$ from $\approx 0.02$ to $\approx 0.08$ for an increase of $x$ from 3.6% to 7.5% was reported in Ref. 36. On the contrary, in Ref. 48 values of $\alpha$ from 0.06 to 0.19 – without any apparent doping trend – were observed for $x$ from 2% to 11%.

For numerical modeling of the measured MO data, we first computed from the LLG equation (Eqs. (A1) and (A2) in the Appendix with the measured magnetic anisotropy fields) the time-dependent deviations of the spherical angles [$\delta\theta(t)$ and $\delta\varphi(t)$] from the corresponding equilibrium values ($\theta_0$, $\varphi_0$). Then we calculated how such changes of $\theta$ and $\varphi$ modify the static magneto-optical response of the sample, which is the signal that we detect experimentally[31]

$$\delta MO(\Delta t, \beta) = -\delta\theta(\Delta t) P^{PKE} + \delta\varphi(\Delta t) P^{MLD} 2\cos 2(\varphi_0 - \beta) + \frac{\delta M_s(\Delta t)}{M_0} P^{MLD} 2\sin 2(\varphi_0 - \beta). \quad (9)$$

The first two terms in Eq. (9) are connected with the out-of-plane and in-plane movement of magnetization, and the last term describes a change of the static magneto-optical response of the sample due to the laser-induced demagnetization.[31] $P^{PKE}$ and $P^{MLD}$ are MO coefficients that describe the MO response of the sample which we measured independently in a static MO experiment,[32,33] and $\beta$ is the probe polarization orientation with respect to the crystallographic direction [100].[31] To further simplify the fitting procedure, we can extract the oscillatory parts from the measured MO data (cf. Fig. 3), which effectively removes the MO signals due to the laser-induced demagnetization [i.e., the last term in Eq. (9)] and due to the in-plane movement of the easy axis [i.e., a part of the MO signal described by the second term in Eq. (9)].[31] Examples of the fitting of the precessional MO optical data are shown in Fig. 7(a) and (b) for (Ga,Mn)As and (Ga,Mn)(As,P), respectively. We stress that in our case the only fitting parameters in the modeling are the damping coefficient $\alpha$ and the initial deviations of the spherical angles from the corresponding equilibrium values. By this numerical modeling we deduced a dependence of the damping factor $\alpha$ on the external magnetic field for two different orientations of $H_{ext}$. At smaller fields, the dependences obtained show a strong anisotropy with respect to the field angle that can be fully ascribed to the field-angle dependence of the precession frequency.[25] However, when plotted as a function of the precession frequency, the dependence on the field-angle disappears – see Fig. 7(c) and (d) for (Ga,Mn)As and (Ga,Mn)(As,P), respectively. For both materials, $\alpha$ initially decreases monotonously with $f$ and finally it saturates at a certain value for $f \geq$ 10 GHz. A frequency-dependent (or magnetic field-dependent) damping parameter was reported in various magnetic materials and a variety of underlying mechanisms responsible for it were suggested as an explanation.[49-51] In our case, the most probable explanation seems to be the one that was used by Walowski et al. to explain the experimental results obtained in thin films of nickel.[49] They argued that in the low field range small magnetization inhomogeneities can be formed – the magnetization does not align parallel in an externally applied field, but forms ripples.[49] Consequently, the measured MO signal which detects sample properties averaged over the laser spot size, which is in our case about 30 μm wide (FWHM), experiences an apparent oscillation damping $\alpha$ because the magnetic properties (i.e., the precession frequencies) are slightly differing within the spot size (see Fig. 6 and 7 in Ref. 49). On the other hand, for stronger external fields the sample is fully homogeneous and, therefore, the precession damping is not dependent on the applied field (the precession frequency), as expected for the intrinsic Gilbert damping coefficient.[52,53] We note that the observed monotonous frequency decrease of $\alpha$ is in fact a signature of a magnetic homogeneity of the studied epilayers.[25] The obtained frequency-independent values of $\alpha$ are $(0.9 \pm 0.2) \times 10^{-2}$ for (Ga,Mn)As and $(1.9 \pm 0.5) \times 10^{-2}$ for (Ga,Mn)(As,P), respectively. The observed enhancement of the magnetization precession damping due to the incorporation of phosphorus is also clearly apparent directly from Figs. 7(a) and 7(b) where the MO data with similar precession frequencies are shown for (Ga,Mn)As and (Ga,Mn)(As,P), respectively. In (Ga,Mn)As the value of $\alpha$ obtained is again fully in accord with the reported Mn doping trend in $\alpha$ in this material.[25] In (Ga,Mn)(As,P), the determined $\alpha$ is similar to the value $1.2 \times 10^{-2}$ which was reported by Cubukcu *et al.* for (Ga,Mn)(As,P) with a similar concentration of Mn and P.[22] Comparing to the doping trends in the series of optimized (Ga,Mn)As materials,[25] the value of $\alpha$ in our (Ga,Mn)(As,P) sample is consistent with the measured Gilbert damping constant in lower Mn-doped (Ga,Mn)As epilayers with similar hole densities and resistivities to those of the (Ga,Mn)(As,P) film.

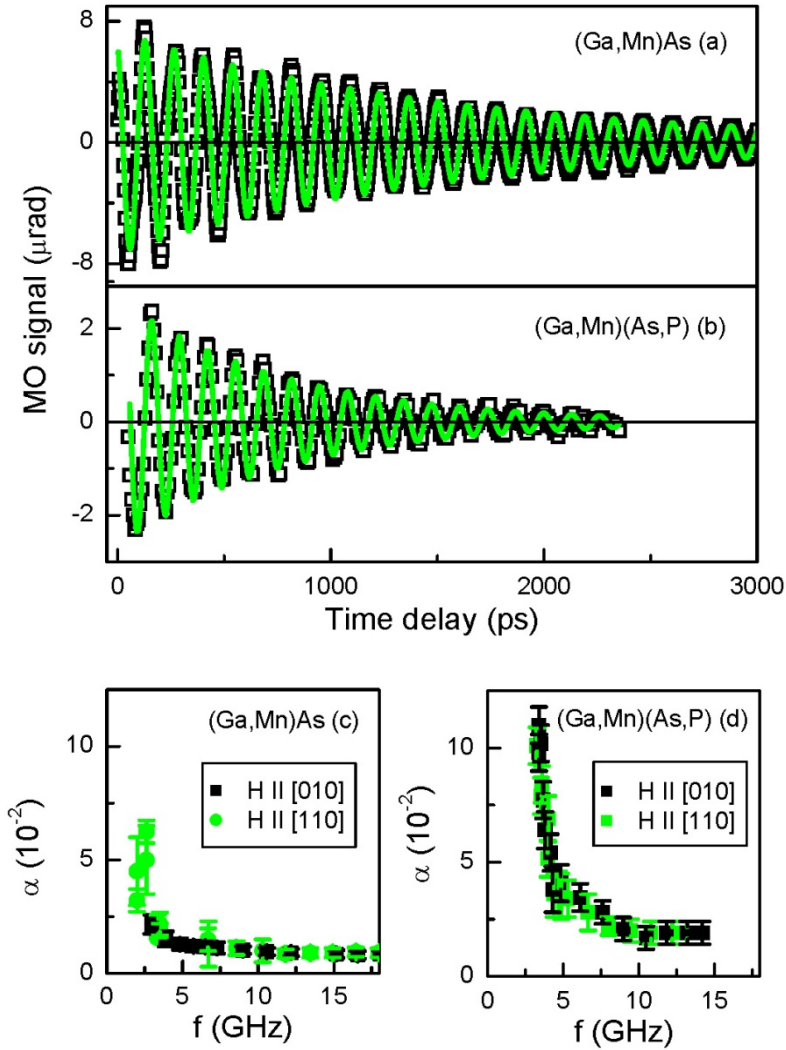

Fig. 7 (Color online): Determination of the Gilbert damping. (a) and (b) Oscillatory part of the MO signal (points) measured in (Ga,Mn)As for the external magnetic field 100 mT (a) and in (Ga,Mn)(As,P) for 350 mT (b); magnetic field applied along the [010] crystallographic direction leads to a similar frequency ($f_0 \approx 7.5$ GHz) in both cases. Lines are fits by the Landau-Lifshert-Gilbert equation. (c) and (d) Dependence of the damping factor ($\alpha$) on the precession frequency for two different orientations of the external magnetic field in (Ga,Mn)As (c) and (Ga,Mn)(As,P) (d).

The high quality of our MO data enables us to evaluate not only the damping of the uniform magnetization precession, which is addressed above, but also the damping of the first SWR. To illustrate this procedure, we show in Fig. 8(a) the MO data measured for $H_{ext} =$ 250 mT applied along the [010] crystallographic direction in (Ga,Mn)As. The experimental data (points) obtained can be fitted by a sum of two exponentially damped cosine functions (line) which enables us to separate, directly in a time domain, the contributions of the individual precession modes to the measured MO signal. In this particular case, the uniform magnetization precession occurs at a frequency $f_0$ = 12.2 GHz and this precession mode is damped with a rate constant $k_{d0}$ = 0.79 ns$^{-1}$. Remarkably, the first SWR, which has a frequency $f_1$ = 23.0 GHz, has a considerably larger damping rate constant $k_{d1}$ = 1.7 ns$^{-1}$ – see Fig. 8(b) where the contribution of individual modes are directly compared and also Fig. 8(c) where Fourier spectra computed from the measured MO data for two different ranges of time delays are shown. To convert the damping rate constant $k_{dn}$ obtained to the damping constant $\alpha_n$ for the *n*-th mode, we can use the generalized analytical solution of the LLG equation. For a sufficiently strong $H_{ext}$ along the [010] crystallographic direction (i.e., when $\varphi \approx \varphi_H = \pi/2$), Eq. (8) can be written as

$$k_{dn} = \alpha_n \frac{g\mu_B}{2\hbar}(2H_{ext} + 2\Delta H_n - 2K_{out} + 2K_C + K_u). \qquad (10)$$

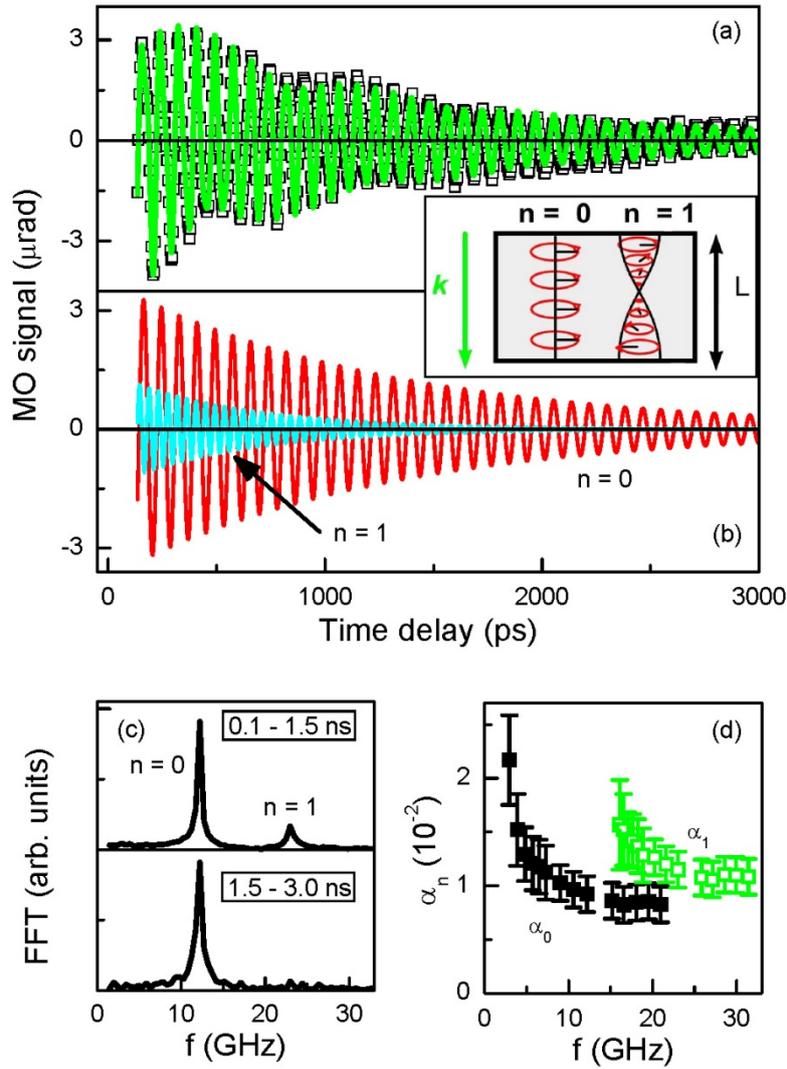

Fig. 8 (Color online): Comparison of the Gilbert damping of the uniform magnetization precession and of the first spin wave resonance. (a) Oscillatory part of the MO signal (points) measured in (Ga,Mn)As for the external magnetic field 250 mT applied along the [010] crystallographic direction. The solid line is a fit by a sum of two exponentially damped cosine functions that are shown in (b). Inset: Schematic illustration[39] of the spin wave resonances with $n$ = 0 (uniform magnetization precession with zero $k$ vector) and $n$ = 1 (perpendicular standing wave with a wave vector $k$ fulfilling the resonant condition $kL = \pi$) in a magnetic film with a thickness $L$. (c) Normalized Fourier spectra computed for the depicted ranges of time delays from the measured MO data, which are shown in (a). (d) Dependence of the damping factor ($\alpha_n$) on the precession frequency for the uniform magnetization precession ($n$ = 0) and the first spin wave resonance ($n$ = 1).

For the case of MO data measured at $H_{ext}$ = 250 mT, the damping constants obtained for modes with *n* = 0 and 1 are $\alpha_0$ = 0.009 and $\alpha_1$ = 0.011, respectively. [We note that the value of $\alpha_0$ obtained from the analytical solution of LLG equation is identical to that determined by the numerical fitting and shown in Fig. 7(c), which confirms the consistency of this procedure.] In Fig. 8(d) we show the dependence of $\alpha_0$ and $\alpha_1$ on the precession frequency. These data clearly show that even if the modes with *n* = 0 and 1 were oscillating with the same frequency, the SWR mode with *n* = 1 would have a larger damping coefficient. However, for sufficiently high frequencies (i.e., external magnetic fields) the damping of the two modes is nearly equal [see Fig. 8(d)]. This feature can be ascribed to the presence of an extrinsic contribution to the damping coefficient for the SWR modes. The extrinsic damping probably originates from small variations of the sample thickness (< 1 nm) within the laser spot size[54] and/or from the presence of a weak bulk inhomogeneity,[43] which is apparent as small variations of $\Delta H_n$. The frequency spacing and the (Kittel) character of the SWR modes is insensitive to such small variations of $\Delta H_n$ but the resulting frequency variations (see Eq. 5) can still strongly affect the observed damping of the oscillations. For high enough external magnetic fields, the variations of $\Delta H_n$ have a negligible role and the damping of the SWR modes is governed solely by the intrinsic Gilbert damping parameter.

## IV. CONCLUSIONS

We used the optical analog of FMR, which is based on a pump-and-probe magneto-optical technique, for the determination of micromagnetic parameters of (Ga,Mn)As and (Ga,Mn)(As,P) DMS materials. The main advantage of this technique is that it enables us to determine the anisotropy constants, the spin stiffness and the Gilbert damping parameter from a single set of the experimental magneto-optical data measured in films with a thickness of only several tens of nanometers. To address the role of phosphorus incorporation in (Ga,Mn)As, we measured simultaneously properties of (Ga,Mn)As and (Ga,Mn)(As,P) with 6% Mn-doping which were grown under identical conditions in the same MBE laboratory. We have shown that the laser-induced precession of magnetization is closely connected with a magnetic anisotropy of the samples. In particular, in (Ga,Mn)As with in-plane magnetic anisotropy the laser-pulse-induced precession of magnetization was observed even when no external magnetic field was applied. On the contrary, in (Ga,Mn)(As,P) with perpendicular-to-plane magnetic EA the precession of magnetization was observed only when the EA position was destabilized by an external in-plane magnetic field. From the measured magneto-optical data we deduced the anisotropy constants, spin stiffness, and Gilbert damping parameter in both materials. We have shown that the incorporation of 10% of P in (Ga,Mn)As leads not only to the expected sign change of the perpendicular-to-plane anisotropy field but also to a considerable increase of the Gilbert damping which correlates with the increased resistivity and reduced itinerant hole density in the (Ga,Mn)(As,P) material. We also observed a reduction of the spin stiffness consistent with the suppression of $T_c$ upon incorporating P in (Ga,Mn)As. Finally, we found that in small external magnetic fields the damping of the first spin wave resonance is sizably stronger than that of the uniform magnetization precession.

## ACKNOWLEDGEMENTS

This work was supported by the Grant Agency of the Czech Republic grant no. P204/12/0853 and 202/09/H041, by the Grant Agency of Charles University in Prague grant no. 1360313 and SVV-2013-267306, by EU grant ERC Advanced Grant 268066 - 0MSPIN, and by Praemium Academiae of the Academy of Sciences of the Czech Republic, from the Ministry of Education of the Czech Republic Grant No. LM2011026, and from the Czech Science Foundation Grant No. 14-37427G.

## APPENDIX

Due to symmetry reasons, it is convenient to rewrite the LLG equation given by Eq. (1) in spherical coordinates where $M_S$ describes the magnetization magnitude and polar $\theta$ and azimuthal $\varphi$ angles characterize its orientation. We define the perpendicular-to-plane angle $\theta$ (in-plane angle $\varphi$) in such a way that it is counted from the [001] ([100]) crystallographic direction and it is positive when magnetization is tilted towards the [100] ([010]) direction (see inset of Fig. 1 for the coordinate system definition). The time evolution of magnetization is given by[37]

$$\frac{dM_S}{dt} = 0, \tag{A1}$$

$$\frac{d\theta}{dt} = -\frac{\gamma}{(1+\alpha^2)M_S}\left(\alpha \cdot A + \frac{B}{\sin\theta}\right) = \Gamma_\theta(\varphi, \theta), \tag{A2}$$

$$\frac{d\varphi}{dt} = \frac{\gamma}{(1+\alpha^2)M_S \sin\theta}\left(A - \frac{\alpha \cdot B}{\sin\theta}\right) = \Gamma_\varphi(\varphi, \theta), \tag{A3}$$

where $A = dF/d\theta$ and $B = dF/d\varphi$ are the derivatives of the free energy density functional $F$ with respect to $\theta$ and $\varphi$, respectively. We express $F$ in a form[10]

$$F = M_S \left[ \begin{array}{c} K_C \sin^2\theta \left(\frac{1}{4}\sin^2 2\varphi \sin^2\theta + \cos^2\theta\right) - K_{out}\cos^2\theta - \frac{K_u}{2}\sin^2\theta(1 - \sin 2\varphi) - \\ -H_{ext}(\cos\theta\cos\theta_H + \sin\theta\sin\theta_H\cos(\varphi - \varphi_H)) \end{array} \right], \tag{A4}$$

where $K_C$, $K_u$ and $K_{out}$ are the constants that characterize the cubic, uniaxial and out-of-plane magnetic anisotropy fields in (Ga,Mn)As, respectively. $H_{ext}$ is the magnitude of the external magnetic field whose orientation is described by the angles $\theta_H$ and $\varphi_H$, which are again counted from the [001] and [100] crystallographic directions, respectively. For small deviations $\delta\theta$ and $\delta\varphi$ from the equilibrium values $\theta_0$ and $\varphi_0$, the Eqs. (A2) and (A3) can be written in a linear form as

$$\frac{d\theta}{dt} = D_1(\theta - \theta_0) + D_2(\varphi - \varphi_0), \tag{A5}$$

$$\frac{d\varphi}{dt} = D_3(\theta - \theta_0) + D_4(\varphi - \varphi_0), \tag{A6}$$

where

$$D_1 = \frac{d\Gamma_\theta}{d\theta}\bigg|_{\theta=\theta_0,\varphi=\varphi_0}, \quad \text{(A7a)}$$

$$D_2 = \frac{d\Gamma_\theta}{d\varphi}\bigg|_{\theta=\theta_0,\varphi=\varphi_0}, \quad \text{(A7b)}$$

and analogically for $D_3$, $D_4$. The solution of Eqs. (A5) and (A6) is expressed by Eqs. (2) and (3) where the magnetization precession frequency $f$ and the damping rate $k_d$ are given by

$$f = \frac{\sqrt{4(D_1 D_4 - D_2 D_3) - (D_1 + D_4)^2}}{4\pi}, \quad \text{(A8)}$$

$$k_d = -\frac{D_1 + D_4}{2}. \quad \text{(A9)}$$

Eqs. (A8) and (A9) for F in the form (A4) can be simplified when the geometry of our experiment – i.e., the in-plane orientation of the external magnetic field ($\theta_H = \pi/2$) – is taken into account. The equilibrium orientation of magnetization is in the sample plane for (Ga,Mn)As ($\theta_0 = \pi/2$) and the same applies for (Ga,Mn)(As,P) if sufficiently strong external magnetic field (see Fig. 1) is applied ($\theta_0 \approx \theta_H = \pi/2$). In such conditions, the precession frequency $f$ and the damping rate $k_d$ are given by the following equations

$$f = \frac{g\mu_B}{2\pi\hbar(1+\alpha^2)}\sqrt{\begin{array}{c}\left(H_{ext}\cos(\varphi - \varphi_H) - 2K_{out} + \frac{K_C(3+\cos 4\varphi)}{2} + 2K_u \sin^2\left(\varphi - \frac{\pi}{4}\right)\right) \times \\ \times (H_{ext}\cos(\varphi - \varphi_H) + 2K_C \cos 4\varphi - 2K_u \sin 2\varphi) + \\ +\alpha^2 \left\{\begin{array}{c}\left(H_{ext}\cos(\varphi - \varphi_H) - 2K_{out} + \frac{K_C(3+\cos 4\varphi)}{2} + 2K_u \sin^2\left(\varphi - \frac{\pi}{4}\right)\right) \times \\ \times (H_{ext}\cos(\varphi - \varphi_H) + 2K_C \cos 4\varphi - 2K_u \sin 2\varphi) - \\ -\left(H_{ext}\cos(\varphi - \varphi_H) - K_{out} + \frac{K_C(3+5\cos 4\varphi)}{4} + \frac{K_u(1-3\sin 2\varphi)}{2}\right)^2\end{array}\right\}\end{array}} \quad \text{(A10)}$$

$$k_d = \alpha \frac{g\mu_B}{2\hbar(1+\alpha^2)}\left(2H_{ext}\cos(\varphi - \varphi_H) - 2K_{out} + \frac{K_C(3+5\cos 4\varphi)}{2} + K_u(1 - 3\sin 2\varphi)\right). \quad \text{(A11)}$$